# Making Sense of Machine Learning: Integrating Youth's Conceptual, Creative, and Critical Understandings of AI

Luis Morales-Navarro, University of Pennsylvania, luismn@upenn.edu (Co-chair)
Yasmin B. Kafai, University of Pennsylvania, kafai@upenn.edu (Co-chair)
Francisco Castro, New York University, francisco.castro@nyu.edu
William Payne, New York University, william.payne@nyu.edu
Kayla DesPortes, New York University, kayla.desportes@nyu.edu
Daniella DiPaola, Massachusetts Institute of Technology, dipaola@mit.edu
Randi Williams, Massachusetts Institute of Technology, randiw12@mit.edu
Safinah Ali, Massachusetts Institute of Technology, safinah@mit.edu
Cynthia Breazeal, Massachusetts Institute of Technology, cynthiab@media.mit.edu
Clifford Lee, Mills College at Northeastern University, cl.lee@northeastern.edu
Elisabeth Soep, YR Media, lissa.soep@yrmedia.org
Duri Long, Northwestern University, duri@northwestern.edu
Brian Magerko, Georgia Institute of Technology, magerko@gatech.edu
Jaemarie Solyst, Carnegie Mellon University, jsolyst@andrew.cmu.edu
Amy Ogan, Carnegie Mellon University, aeo@andrew.cmu.edu
Cansu Tatar, North Carolina State University, ctatar@ncsu.edu
Shiyan Jiang, North Carolina State University, sjiang24@ncsu.edu
Jie Chao, Concord Consortium, jchao@concord.org
Carolyn P. Rosé, Carnegie Mellon University, cp3a@andrew.cmu.edu
Sepehr Vakil, Northwestern University, sepehr.vakil@northwestern.edu (Discussant)

**Abstract:** Understanding how youth make sense of machine learning and how learning about machine learning can be supported in and out of school is more relevant than ever before as young people interact with machine learning powered applications everyday—while connecting with friends, listening to music, playing games, or attending school. In this symposium, we present different perspectives on understanding how learners make sense of machine learning in their everyday lives, how sensemaking of machine learning can be supported in and out of school through the construction of applications, and how youth critically evaluate machine learning powered systems. We discuss how sensemaking of machine learning applications involves the development and integration of conceptual, creative, and critical understandings that are increasingly important to prepare youth to participate in the world.

## Symposium Overview

Recent calls to promote artificial intelligence (AI) literacy in K-12 education highlight the importance of engaging young learners with big ideas, preparing them for careers in computing and to be critical consumers and designers of technology (Touretzky et al., 2019; DiPaola et al., 2022; Wang et al., 2021). Within AI literacy, fostering an understanding of machine learning (ML), which involves the use of data rather than code to shape the behavior of computer programs, is crucial (Long & Magerko, 2020; Zimmermann-Niefield et al., 2019). Machine learning is a new paradigm in computing education (Shapiro & Tissenbaum, 2019) that learners must engage with to become computationally literate and be empowered to participate in computing (Kafai & Proctor, 2022).

Despite the fact that machine learning is often black boxed in consumer applications, research shows that children construct naive explanations to make sense of how these work (Druga et al., 2017; Williams et al., 2019). At the same time, designing machine learning applications requires thinking like a scientist and building hypotheses, using data sets to train and test models to make predictions (Shapiro et al., 2018; Langley, 1988). Both using and building ML-powered applications demand making sense of how models work and how data shape their behaviors. Yet little attention has been given to how youth integrate conceptual, critical, and creative understandings in making sense of ML-powered applications. Sensemaking involves explaining observed phenomena using theory and evidence (Newman et al., 1993; Crowder, 1996) to "figure something out" by dynamically building and revising explanations using both formal and everyday knowledge (Odden & Russ, 2018). Whereas sensemaking has traditionally centered on conceptual understanding, learners also engage in explanation building when considering ethics to critically understand how systems work. At the same time, making applications engages learners in creative understanding, by having to make decisions on how to create personally relevant projects they build and revise explanations. In this symposium we bring together research on young people's novice sensemaking of ML, that is how they come up with and revise their explanations of how machine learning works as well as how their sensemaking can be supported through the analysis and construction of ML-powered creative applications in and out of school. Participants discuss the following questions:

    (a) How do youth build on their everyday experiences with technology to make sense of ML?

    (b) How can youth sensemaking of ML be supported through the design of ML-powered applications?

    (c) How do youth critically evaluate and understand ML-powered systems?

The invited works provide examples of how conceptual, creative, and critical understandings of ML and AI can be integrated. Presenters apply Learning Sciences perspectives on embodied cognition, critical literacies, modeling and design-based research to the analysis of ML sensemaking with studies conducted in computing and non-STEM contexts, out of school and in K-12 classrooms: Castro and colleagues investigate youth's understanding and embodied learning of ML and in a computing and dance intervention; DiPaola and colleagues examine how embedding ethics into three project-based curricula supported students to develop understandings of ML applications as sociotechnical systems; Lee and Soep study meaning making processes when youth create projects about and with AI/ML technologies through a critical perspective; Long and Magerko research how embodied interaction through dance can support learning about ML in informal spaces; Morales-Navarro and Kafai investigate how youth make sense of ML when encountering failure cases as users and creators of applications; Solyst and Ogan study girl's funds of knowledge and knowledge gaps around AI/ML and fairness; Tatar and colleagues adopt a situated learning perspective to analyze students' data modeling experiences and their impact on shaping students' understanding of AI/ ML.

The symposium is organized in three sections: (1) the chairs will introduce the topic and then each presenter will give an one-minute teaser about their work (~10 min); (2) the first half of the presenters will have 20 minutes to share their work using posters placed around the room, followed by the second half of presenters (20 minutes)—this arrangement will give the audience and presenters time to see each other's posters; (3) our discussant Sepehr Vakil, an expert in justice-centered computing with experience in community-centered AI education, will synthesize and reflect on findings (10 minutes) followed by a Q&A with audience and presenters (~15 min).

# 1. Shuttling Between Contextualized Creative Computing and Learners' Understanding of Machine Learning Algorithms in the Real-World

Francisco Castro, New York University, francisco.castro@nyu.edu

William Payne, New York University, william.payne@nyu.edu

Kayla DesPortes, New York University, kayla.desportes@nyu.edu

Integrating movement practices from dance into computing and machine learning (ML) education can lead to culturally sustaining experiences where learners draw upon cultural ways of knowing as they explore identity across individual, social, and political dimensions (Castro et al., 2022; DesPortes et al., 2022; Payne et al., 2021). Novel dance-computing technologies and curricula, like danceON, must not only leverage cultural assets but set out to sustain communities (Paris & Alim, 2014). In our work, we explore how BIPOC learners shuttle between embodied knowledge and reasoning of ML concepts and how their personal and sociopolitical understanding of ML reaches beyond the learning environment.

Wild and Pfannkuch (1999) articulate the importance of learners shuttling between their disciplinary and real-world knowledge in statistical thinking. Researchers have found this can support purposeful and meaningful exploration of concepts (Ben-Zvi & Aridor-Berger, 2016). We draw on their framing of shuttling as we examine how learners build their mental models of computing systems through creating dance-computing artifacts, then leverage this understanding to explore personal experiences and implications of computing systems in society.

We examined a section of a 15-week internship with 6 high school learners from STEM From Dance (stemfromdance.org), a community organization that engages young women of color in STEM and dance. Within these sessions, learners trained classifiers to identify two poses with Google's Teachable Machine while also working with danceON, a creative coding environment that enables learners to code animations over dance videos (Payne et al., 2021). In tandem, they read articles about misuse and biases within ML systems. We reviewed two sessions that were recorded and transcribed verbatim and identified instances where learners discussed general ML concepts (who uses and develops systems), ML processes (training, testing, etc.), and ML behaviors (pose detection); this enabled us to identify when learners (1) reasoned about ML concepts with their body and (2) shuttled between their embodied experience of ML, their personal experiences of ML, and understanding of ML in society.

Learners regularly encountered limitations of pose detection as they witnessed animations and body points drawn incorrectly. They hypothesized possible causes such as occlusion, movement speed, and clothing, and incorporated strategies for improving accuracy through movement. Through making classifiers with Teachable Machine, learners identified gaps in the training data and discussed how more iterations of data collection may improve accuracy. Finally, in discussions, learners connected their conceptions and understanding of pose detection with personal experiences with other ML systems, such as limited performance of home assistants within a family of non-native English speakers and reasoned about the impact of biased ML systems for others. Our work highlights the affordances of creative dance computing spaces as avenues toward the embodied learning of ML. By engaging in activities and discussions of ML within the embodied, cultural, and collaborative nature of dance, we found concrete instances of shuttling between their understanding built through dance and their understanding of the real-world context—i.e., the impact of human-driven design decisions on the implementation of ML, and their personal experiences with AI systems. Through continued exploration of co-designed scaffolds, we can further develop the ways in which asset-based experience in cultural practices like dance can facilitate sociopolitical examination of ML, which is a key component of supporting culturally sustaining pedagogy.

## 2. Use, Understand, Create: Embedding Ethics in Machine Learning Curricula for Middle School Youth

Daniella DiPaola, Massachusetts Institute of Technology, dipaola@mit.edu
Randi Williams, Massachusetts Institute of Technology, randiw12@mit.edu
Safinah Ali, Massachusetts Institute of Technology, safinah@mit.edu
Cynthia Breazeal, Massachusetts Institute of Technology, cynthiab@media.mit.edu

Machine Learning (ML) is a powerful computational tool that can greatly benefit, but also potentially harm users, especially those from systematically marginalized communities (Buolamwini & Gebru 2018, Noble 2018, O'Neil 2016). The tension between benefits and harms should be presented in an age-appropriate way to all students learning about ML. However, recent work shows that computer science educators typically withhold or exclude ethical issues in their courses, to the detriment of their students (Fiesler, Garrett, & Beard, 2020).

This work discusses the design principle "embedded ethics" and how educators can incorporate ethical thinking into project-based ML curricula. "Embedded ethics" involves teaching ethics alongside technical concepts helps students develop a fuller understanding of the technology, including the long-term implications of systems they create (Saltz et al., 2019, Skirpan et al. 2018, DiPaola, Payne, & Breazeal 2020). This work aligns with the Use-Understand-Create framework for digital literacy and shows how ethical thinking can be a part of every stage of students' learning progression (MediaSmarts, 2021).

This paper presents three project-based curricula on emerging ML topics – generative adversarial networks, affective perception, and supervised machine learning (Williams et al., 2022). Each curriculum includes the use of real-world ML demos, conceptual discussion, and open-ended creation to teach students about ethical thinking. In an introductory activity in Creative AI, students learn to anticipate potential beneficial and harmful uses of various generative AI tools. In Dancing With AI, students employ their technical understanding to predict the impact of ML classifiers trained with flawed datasets. In How to Train Your Robot, students consider different stakeholders' values in their final project designs. In the summer of 2020, we trained 11 middle school teachers, primarily from Title 1 schools, to co-teach one of the three curricula to 78 middle-school students from 8 states across the USA. A mixed methods approach was used to evaluate their mastery of ethical thinking in different activities. This encompassed statistical methods to evaluate AI concept surveys pre-post, thematic coding to evaluate classwork, and rubrics to evaluate final projects. These measures were developed by the authors for the purposes of this project.

Before the workshops, the surveys showed that most students had an agreeable disposition toward AI, associating it with positive words like "exciting" and positive impacts like "making jobs easier." Students' classwork demonstrated an increased ability to imagine potential societal repercussions of AI systems and apply ethical decision making to their final projects. In students' final project creations, students were able to transfer their knowledge of ethical implications of AI systems to areas of personal interest. Overall, the authors observed that embedding ethics into curricula led to students developing a nuanced understanding of ML applications as sociotechnical systems.

This paper describes three approaches taken to teach students about ethical thinking throughout all stages of their learning trajectory. In previous work, researchers found that tools for ethical analysis enabled students to critique and then redesign AI systems that they were familiar with (DiPaola, Payne, & Breazeal 2020). The three empirical studies examined in this work showed students going one step further and implementing reimagined AI systems through their own projects. For example, one student in the Dancing with AI curriculum, inspired by the effects of COVID-19 on their community, created a project to classify different types of masks based on their effectiveness.

## 3. Learning with and about Ethical Artificial Intelligence through Youth-Made Media

Clifford Lee, Mills College at Northeastern University, cl.lee@northeastern.edu
Elisabeth Soep, YR Media, lissa.soep@yrmedia.org

This study seeks to understand how young people underrepresented in STEM make meaning of the role of AI in their lives and society and how their relationship to the technology evolves when they create their own AI-based tools and media. This research analyzes the curriculum and pedagogy behind three ethics-centered AI learning activities housed within an after-school multimedia production organization. Critical Computational Expression (CCE) is a conceptual and pedagogical framework that integrates the three distinct traditions of: critical pedagogy, computational thinking, and creative expression (Lee & Soep, 2022). Interactive youth developers are attuned to the aesthetics, design, and creative representations of their products while being conscious of the sociopolitical messages and computational sophistication of their interactive stories (Lee & Soep, 2016).

Our ethnographic study centered participant observation, through audio recordings of moment-to moment interactions in class, end-of-session focus group interviews, and analyses of youth-generated artifacts within the learning environment over the course of two and a half years. Our research team used a grounded theory approach to code, reduce, and analyze the data to generate themes according to our Critical Computational Expression framework. This paper addresses the following research questions:

1. What can we learn about young people's understanding of AI when they produce media with and about it?

2. What are the design features of an ethics-centered pedagogy that promotes STEM engagement via AI? In the three activities examined in this study, young people: repurposed their phones' text autocomplete features to produce poetry; countered Spotify's system for rating pop songs' danceability by designing an interactive experience of their own; and developed a drawing tool inviting users to scribble over photographs of faces to determine what degree of disguise was required to dodge facial recognition software. In total, sixteen producers, aged fifteen to twenty-four, who are predominantly youth of color and those contending with economic and other barriers to full participation in STEM fields, engaged in this study.

Our findings suggest that students can feel disempowered by their increasingly intelligent technologies. Through ongoing observation and analysis, we see students deepen curiosity about and understanding of AI that allows them to exercise agency and conceptualize creative projects using their new knowledge to manipulate and hack AI-dependent algorithms. Additionally, our results show that participating in ethics-centered learning activities and developing AI-powered tools do not create a permanent evolution of youth's understanding of their agency as it's related to AI in their lives; instead, these modes of involvement offer meaningful glimpses into how the problematic dimensions of AI systems are pervasive, yet not undefeatable in terms of young people's positioning with respect to technology and their role in the culture it produces.

By drawing on digital tools and practices that youth are familiar with as consumers, young people develop sufficient technical know-how, creative engagement, and critical curiosity about the implications of these systems to demystify how everyday tools work, then start envisioning ways to spark new action and conversation. Understanding the mechanisms that shape human interactions with AI to conform with the patterns embedded in its functioning afforded students the opportunity to discover ways to disrupt these systems with creativity, originality, and new ways of thinking.

## 4. Using Embodied Interaction and Creative Making to Foster Machine Learning Sensemaking in Informal Learning Contexts

Duri Long, Northwestern University, duri@northwestern.edu
Brian Magerko, Georgia Institute of Technology, magerko@gatech.edu

Our research explores how embodied interaction and creative making can engage family groups in discussions surrounding machine learning in informal learning contexts like homes and museums. We design activities to support creative, embodied learning experiences and study participants' learning talk and interest development surrounding these activities.

We conceptualize embodied interaction as physical interaction with and/or control of the activity. Creative making refers to the production of personally relevant artifacts, especially those that persist beyond the activity. We hypothesize that the emphasis of these constructs on the self can help learners reconceptualize AI and ML as relevant areas of interest for people "like me" (Papastergiou et al., 2008; Magerko et al., 2016; Guzdial et al., 2013; Buechley et al., 2008). In addition, we hypothesize based on prior work (e.g., Antle et al., 2013; Horn et al., 2009; Sulmont 2019) that embodied interaction can make abstract concepts–like AI–concrete for learners.

We have engaged in design research to develop two activities that support learning about ML. The first, *Creature Features*, engages learners in building a training dataset for a feature-based ML bird classification algorithm using a tangible interface. Learners can explore issues like dataset bias and representation using cards and tokens and can iteratively revise their datasets after viewing the algorithm's results. The second activity, *LuminAI*, engages learners in improvising dance with an AI partner. After dancing with the AI, learners can explore an interactive, 3D visual representation of the way the dancer uses unsupervised ML to group gestures in memory. We recruited 14 family groups (38 participants; 21 age 6-17 and 17 age 18+) to interact with the prototypes in their homes. Family members were given an age-appropriate survey following their interaction. We asked both Likert-scale and free-response questions to assess interest development, content knowledge gain, and whether the activities elicited creative interactions. We also had family members record audio of their interactions with the activities. We coded the transcribed dialogue to identify instances of *learning talk*—i.e., conversation that was relevant to the learning goals of the activity (Roberts & Lyons, 2017). The quantitative results from the survey data supplemented with the qualitative analysis of the audio transcripts provided insight into which exhibits led to learning talk, interest development in AI, content knowledge gain, and creative engagement.

The tangible interface and iterative cycle of testing and revision in *Creature Features* supported in-depth discussion of features and their impact on the algorithm. The activity was most successful with families with kids aged 10+, and more scaffolding may be needed to help learners connect the activity with "real-world" technologies and issues. Although *LuminAI* focused on teaching AI through a creative activity, the ephemeral nature of dance means learners did not generate a lasting artifact. This may have limited the impact of *LuminAI* on interest development. Although learners scored well on the content knowledge questions related to unsupervised ML in *LuminAI*, learners expressed that they were intimidated by the interface. This suggests that interactive visual interfaces–even when building on embodied metaphors in a creative domain–may necessitate additional scaffolding for novice audiences. Our results indicate that tangible interaction can be an effective design feature for promoting sensemaking about ML. Future research is needed to examine whether creative making can be an effective design feature for promoting learning about ML. Our work contributes to understanding how to design casual AI learning experiences for novices that can integrate into everyday life.

## 5. Youth's Sensemaking through Failure Cases in Machine Learning Powered Applications

Luis Morales-Navarro, University of Pennsylvania, luismn@upenn.edu
Yasmin B. Kafai, University of Pennsylvania, kafai@upenn.edu

Youth encounter machine learning (ML) applications every day and while several studies have investigated their understanding of how machine learning works (Druga et al., 2017; Williams et al., 2019) most of it has centered around success, that is when ML applications work as expected. We present results from an exploratory study in which we investigate how youth make sense of ML when encountering failure cases as users and creators of physical computing applications. By investigating how youth make sense of failure cases we aim to analyze youth's conceptual understanding as well as their consideration of the limitations and implications of ML technologies.

Whereas conversations about ML, society and ethics are often disconnected from technical issues (Fiesler, 2020; Petrozzino, 2021), the consequences and implications of ML applications are closely intertwined with functionality failures (Raji et al., 2022). To investigate youth's sensemaking of ML and its implications, that is how they build and revise explanations using both formal and everyday knowledge (Odden & Russ, 2018), we build on previous work on youth's interaction with failure artifacts. Failure artifacts are applications that have deliberate failures, bugs or mistakes and can elicit learners' understanding of how computing applications work (Fields et al., 2021). At the same time, encountering failure when creating projects can support sensemaking as students resolve failure cases, avoid recurring failure, prepare for novel failures, and calibrate their confidence (DeLiema et al., 2022).

We conducted a ML+eTextiles workshop at a science center in the Northeastern United States with 12 (15-16 years old) youths of Color during the Spring and Summer 2022. In the Spring, youths were presented with consumer applications with failure cases and eTextiles ML-powered failure artifacts. Using stickies and big paper methods (Yip et al., 2013, Woodward, 2018), we asked them to brainstorm how these applications worked and if they encountered any failures how they could fix them. Following, during the summer, youths learned to create ML-powered eTextiles and designed their own personally relevant projects. As they worked on their projects we had several sessions during which they reflected on failure cases they encountered. We analyzed workshop artifacts and recordings using iterative thematic analysis (Braun and Clarke, 2012).

Prior to any instruction on ML, half of the youth voiced ideas of how ML applications used data to generate predictions. They also brought up issues of bias and its ethical implications when discussing failure cases in consumer technologies and reflected on their personal experiences using some of these technologies. As they built models for their eTextile projects, they encountered failure cases particularly with regards to diversity of training data and overfitting. These failure cases generated discussions on the importance of testing and iterative design of data sets, considering who and how the projects would be used, and anticipating how the projects would affect people. This poster provides evidence of youth's understanding of ML as users and producers when encountering failure cases. Using failure cases and failure artifacts in instruction and reflection on failure in their own creations may be particularly helpful to foster youth's understanding of how technical failure and social implications and limitations may be intertwined.

## 6. Talking about Fairness in Artificial Intelligence and Machine Learning with Girls

Jaemarie Solyst, Carnegie Mellon University, jsolyst@andrew.cmu.edu
Amy Ogan, Carnegie Mellon University, aeo@andrew.cmu.edu

Children are the future users and creators of AI. However, girls and particularly girls of color are often excluded in the design of technology while simultaneously being greatly impacted by issues of fairness in AI. Educational opportunities about AI and ethics designed for girls, their interests, and their funds of knowledge, are essential. Through a series of workshops, we aimed to understand those funds of knowledge as well as their sense making, including knowledge gaps around AI and fairness, focusing on the critical age prior to high school.

In our workshops, we based our learning materials on culturally responsive computing (CRC). Positioning learners as technosocial change agents, poised to advocate for justice in technology, CRC leverages: Asset-building by adding onto what learners know, Reflection by prompting learners to critically analyze and decompose existing power structures, and Connectedness by strengthening and taking into account relationships that the learners have within the learning environment and their broader communities (Scott et al., 2015).

We ran a series of six workshops online and in-person with middle school girls (11-14) and an additional in-person workshop with fifth and sixth graders (ages 9-12). Workshop material was based on CRC with a main focus on AI. We introduced the basics (e.g., what is an algorithm, what is AI, and how training data is involved in machine learning). We then instigated discussions around various AI technologies and how they could be fairer, as well as design activities where learners made sense of and thought of their own AI-powered inventions. We conducted thematic analysis on transcripts, chat logs, and artifacts.

Our findings showed that girls' ideas around fairness followed models of equality/inequality and nice/kind vs. rude/mean, i.e., many learners defined fairness as everyone getting the same resources, as well as having kind interactions (e.g., a nice robot), while unfairness was defined as the opposite. Some learners brought up more complex ideas like equity, or technology accommodating user differences. We saw that in older groups (e.g., seventh and eighth graders), learners had existing understandings of bias and could apply it to technology being unfair. Younger girls and those with lower prior knowledge in computing needed more scaffolding to critically discuss and question technology. Middle school learners were able to see unfairness in existing examples of AI (e.g., bias in Google search image results). Lastly, applying fairness to more technical aspects of AI, such as how training data can impact ethical ML, was a more challenging topic that needed more time and explanation for learners who did not have a lot of prior STEM exposure.

This suggests that fairness should be a main focus of AI and ethics, supporting exposure to more complex ideas about fairness (e.g., equality vs. equity) before considering fairness applied to more technical topics, such as training data. It also indicates that this is a critical set of ages at which conceptions of fairness are beginning to emerge and shift; perhaps necessitating different educational interventions, where for younger children, fairness and ethics may be a specific topic. Older children (e.g., 12-14) were more able and eager to discuss power and privilege, but younger children may need considerably more scaffolding for these topics in a curriculum.

# 7. The Impact of a Technology-Enhanced Unit on High School Students' Understanding of Artificial Intelligence & Machine Learning

Cansu Tatar, North Carolina State University, ctatar@ncsu.edu
Shiyan Jiang, North Carolina State University, sjiang24@ncsu.edu
Jie Chao, Concord Consortium, jchao@concord.org
Carolyn P. Rosé, Carnegie Mellon University, cp3a@andrew.cmu.edu

Despite the substantial interest in K-12 AI curriculum development, there is a lack of intervention-based research, in particular regarding the effects of AI curriculum on K-12 students' understanding of AI and machine learning (Chiu et al., 2021; Estevez et al., 2019). This study explores high school students' understanding of AI & ML before and after a technology-enhanced curriculum intervention. We adopt a situated learning perspective as our theoretical framework to understand students' data modeling experiences and their impact on shaping students' understanding of AI & ML. Modeling has been an effective learning strategy for knowledge construction that describes a process of developing representations of phenomena being experienced in order to engender conceptual change (Jonassen, 2011).

A Journalism teacher participated in our professional development workshop for four weeks and implemented our technology-enhanced AI curriculum in her Journalism classroom for three weeks. This class included twenty-eight students: three 10th graders, nine 11th graders, and sixteen 12th graders. Students used StoryQ to build machine learning models with text data. Students did not receive any formal training in AI before this curriculum intervention. Before and after the curriculum intervention, students completed a knowledge assessment. This assessment was designed by following the scenario-based representation model (Sun et al., 2003) and validated by machine learning experts and teachers. Students' responses were analyzed by following open-coding strategies (Strauss & Corbin, 1994), focusing on their understanding of AI & ML and reasoning about everyday AI technology.

Our analyses revealed that before the curriculum intervention, students mostly viewed AI as robots that help people perform certain tasks. This could be due to the media portrayal of AI as robots, which is one of the most common themes in movies and popular culture. After the curriculum intervention, students defined AI as a technology that mimics human intelligence. We also asked students to share their understanding of ML. In the pre-assessment, most students indicated that they were not familiar with the term. Their responses included random guesses like learning through machines. After the intervention, their conceptions shifted from random guesses to more detailed explanations including feature selection and recognizing the importance of training data for model development. Additionally, we found that students mostly viewed Siri as a pre-programmed technology before curriculum implementation. In the post-assessment, most students used AI concepts to explain the working mechanism of Siri. During the curriculum intervention, students experienced feature selection when building ML models. This hands-on experience in building ML models might help them reason about how virtual assistants (e.g., Siri or Alexa) work. This study demonstrated that a technology-enhanced AI curriculum offering opportunities for building ML models helped high school students gain a more in-depth understanding of AI & ML and its applications. A fertile area for future studies is exploring patterns of model development in different kinds of ML modeling tasks and investigating how to best support students' diverse ways of building models.

## Acknowledgements

Work for Shuttling Between Contextualized Creative Computing and Learners' Understanding of Machine Learning Algorithms in the Real-World was supported by the National Science Foundation (NSF) under Grant # 2127309 to the Computing Research Association for the CIFellows 2021 Project and STEM+C 1933961. Work for Learning with and about Ethical Artificial Intelligence through Youth-Made Media was supported by a grant from the NSF, Division of Research on Learning in Formal and Informal Settings to YR Media (AISL Award #1906895 and DRL #1647150-AM0005, subaward number UWSC9758). Work for "Using Embodied Interaction and Creative Making to Foster Machine Learning Sensemaking in Informal Learning Contexts" was supported by a grant from the NSF, Division of Research on Learning in Formal and Informal Settings (DRL-2214463) to Georgia Institute of Technology. Work for "Use, Understand, Create: Embedding Ethics in Machine Learning Curricula for Middle School Youth" was supported by grants from Amazon Future Engineer, NSF Graduate Research Fellowship Program #1745302, and the LEGO Foundation. We are also grateful to I2 Learning, the MIT STEP Lab, and our research interns Tejal Reddy, Victor Sindato, Grace Kim, and Ryan Blumofe for their contributions to the development and instruction of the summer workshops. Work for The Impact of a Technology-Enhanced Unit on High School Students' Understanding of Artificial Intelligence & Machine Learning was supported by a grant from the NSF to Concord Consortium, North Carolina State

University, and Carneige Mellon University (DRL-1949110). Morales-Navarro & Kafai are grateful to Mia Shaw and Yilin Liu for their support in data collection for "Youth's Sensemaking through Failure Cases in Machine Learning Powered Applications". Any opinions, findings and conclusions or recommendations expressed in this paper are those of the authors and do not necessarily reflect the views of the funding agencies, the University of Pennsylvania, New York University, Massachusetts Institute of Technology, Northeastern University, YR Media, Northwestern University, Georgia Institute of Technology, Carnegie Mellon University, North Carolina State University, or Concord Consortium.

# References

Ali, S., Payne, B. H., Williams, R., Park, H. W., & Breazeal, C. (2019, June). Constructionism, ethics, and creativity: Developing primary and middle school artificial intelligence education. In *International workshop on education in artificial intelligence k-12 (eduai'19)* (pp. 1-4)

Antle, A. N., Corness, G., & Bevans, A. (2013). Balancing justice: Comparing whole body and controller-based interaction for an abstract domain. *International Journal of Arts and Technology*, 6(4), 388-409.

Ben-Zvi, D., & Aridor-Berger, K. (2016). Children's wonder how to wander between data and context. In *The teaching and learning of statistics* (pp. 25-36). Springer, Cham.

Braun, V., & Clarke, V. (2012). Thematic analysis. In H. Cooper, P. M. Camic, D. L. Long, A. T. Panter, D. Rindskopf, & K. J. Sher (Eds.), *APA handbook of research methods in psychology, Vol. 2. Research designs: Quantitative, qualitative, neuropsychological, and biological* (pp. 57–71). American Psychological Association.

Buechley, L., Eisenberg, M., Catchen, J., & Crockett, A. (2008, April). The LilyPad Arduino: using computational textiles to investigate engagement, aesthetics, and diversity in computer science education. In *Proceedings of the SIGCHI conference on Human factors in computing systems* (pp. 423-432).

Buolamwini, J., & Gebru, T. (2018, January). Gender shades: Intersectional accuracy disparities in commercial gender classification. In Conference on fairness, accountability and transparency (pp. 77-91). PMLR.

Castro, F. E. V., DesPortes, K., Payne, W., Bergner, Y., & McDermott, K. (2022). AI + Dance: Co-Designing Culturally Sustaining Curricular Resources for AI and Ethics Education Through Artistic Computing. Proceedings of the 2022 ACM Conference on International Computing Education Research - Volume 2, 26–27. https://doi.org/10.1145/3501709.3544275

Chiu, T. K. F., Meng, H., Chai, C. S., King, I., Wong, S., & Yam, Y. (2021). Creation and evaluation of a pre-tertiary artificial intelligence (ai) curriculum. *IEEE Transactions on Education*.

Crowder, E. M. (1996). Gestures at work in sense-making science talk. *The Journal of the Learning Sciences, 5*(3), 173-208.

DeLiema, D., Kwon, Y. A., Chisholm, A., Williams, I., Dahn, M., Flood, V. J., ... & Steen, F. F. (2022). A Multi-dimensional Framework for Documenting Students' Heterogeneous Experiences with Programming Bugs. *Cognition and Instruction*, 1-43.

DesPortes, K., McDermott, K., Bergner, Y., & Payne, W. (2022). "Go[ing] hard...as a woman of color": A case study examining identity work within a performative dance and computing learning environment. *ACM Transactions on Computing Education*. https://doi.org/10.1145/3531000.

DiPaola, D., Payne, B. H., & Breazeal, C. (2020, June). Decoding design agendas: an ethical design activity for middle school students. In *Proceedings of the interaction design and children conference* (pp. 1-10).

DiPaola, D., Payne, B. H., & Breazeal, C. (2022) Preparing Children To Be Conscientious Consumers And Designers Of Ai Technologies. In S. Kong, and H. Abelson, (Eds.), *Computational Thinking Education in K-12: Artificial Intelligence Literacy and Physical Computing*. MIT Press.

Druga, S., Williams, R., Breazeal, C., & Resnick, M. (2017, June). " Hey Google is it ok if I eat you?" Initial explorations in child-agent interaction. In *Proceedings of the 2017 conference on Interaction Design and Children* (pp. 595-600).

Druga S., & Ko J A. 2021. How do children's perceptions of machine intelligence change when training and coding smart programs?. In *Interaction Design and Children* ACM, Athens Greece, 49–61.

Estevez, J., Garate, G., & Graña, M. (2019). Gentle introduction to artificial intelligence for high-school students using scratch. IEEE access, 7, 179027-179036.

Fields, D. A., Kafai, Y. B., Morales-Navarro, L., & Walker, J. T. (2021). Debugging by design: A constructionist approach to high school students' crafting and coding of electronic textiles as failure artefacts. *British Journal of Educational Technology, 52*(3), 1078-1092.

Fiesler, C., & Garrett, N. (2020). Ethical Tech Starts With Addressing Ethical Debt. *Wired.*

Fiesler, C., Garrett, N., & Beard, N. (2020, February). What do we teach when we teach tech ethics? a syllabi analysis. In *Proceedings of the 51st ACM Technical Symposium on Computer Science Education* (pp. 289-295).

Guzdial, M. (2013). Exploring Hypotheses about Media Computation. *Proceedings of the Ninth Annual International ACM Conference on International Computing Education Research*. (pp.19–26).

Horn, M. S., Solovey, E. T., Crouser, R. J., & Jacob, R. J. (2009, April). Comparing the use of tangible and graphical programming languages for informal science education. In *Proceedings of the SIGCHI Conference on Human Factors in Computing Systems* (pp. 975-984).

Jiang, S., DesPortes, K., Bergner, Y., Zhang, H., Lee, I., Moore, K., Cheng, Y., Perret, B., Walsh, B., Guggenheim, A., Dalton, B., Forsyth, S., Yeh, T., Akram, B., Yoder, S., Finzer, W., Chao, J., Rosé, C. P., Payne, W., Castro-Norwood, F., McDermott, K. (2022) Agents, Models, and Ethics: Importance of Interdisciplinary Explorations in AI Education. In *Proceedings of the International Society of the Learning Sciences*.

Kafai, Y. B., & Proctor, C. (2022). A Revaluation of Computational Thinking in K–12 Education: Moving Toward Computational Literacies. *Educational Researcher, 51*(2), 146-151.

Jonassen, D. H. (2011). *Learning to solve problems: A handbook for designing problem-solving learning environments* (1st ed.). Routledge.

Langley, P. (1988). Machine learning as an experimental science. *Machine Learning, 3*(1), 5-8.

Lee, C. & Soep , E. (2016) None But Ourselves Can Free Our Minds: Critical Computational Literacy as a Pedagogy of Resistance. *Equity & Excellence in Education, 49*(4), 480-492, DOI: 10.1080/10665684.2016.1227157

Lee, C., & Soep, E. (2022). Code for What?: Computer Science for Storytelling and Social Justice. *MIT Press*.

Long, D., & Magerko, B. (2020, April). What is AI literacy? Competencies and design considerations. In *Proceedings of the 2020 CHI conference on human factors in computing systems* (pp. 1-16).

Magerko B, Freeman J, McKlin T, Reilly M, Livingston E, McCoid S, Crews-Brown A (2016). EarSketch: A STEAM-Based Approach for Underrepresented Populations in High School Computer Science Education. *ACM Transactions on Computing Education (TOCE). 16*(4):14. https://doi.org/10.1145/2886418

MediaSmarts. (2021, August 27). Digital Literacy Fundamentals. Retrieved July 20, 2022, from https://mediasmarts.ca/digital-media-literacy/general-information/digital-media-literacy-fundamentals/digital-literacy-fundamentals

Newman, D., Morrison, D., & Torzs, F. (1993). The world in the classroom: Sense-making and seasonal change. *Interactive Learning Environments, 3*, 1-16.

Noble, S. U. (2018). *Algorithms of oppression*. New York University Press.

O'Neil, C. (2016). Weapons of math destruction: How big data increases inequality and threatens democracy. Broadway books.

Odden, T. O. B., & Russ, R. S. (2019). Defining sensemaking: Bringing clarity to a fragmented theoretical construct. *Science Education, 103*(1), 187-205.

Papastergiou, M. (2008). Are computer science and information technology still masculine fields? High school students' perceptions and career choices. *Computers and Educatio. 51*(2):594–608.

Paris, D., & Alim, H. S. (2014). What are we seeking to sustain through culturally sustaining pedagogy? A loving critique forward. *Harvard Educational Review*, 84(1), 85-100.

Payne, W. C., Bergner, Y., West, M. E., Charp, C., Shapiro, R. B. B., Szafir, D. A., Taylor, E. V., & DesPortes, K. (2021). danceON: Culturally Responsive Creative Computing. *Proceedings of the 2021 CHI Conference on Human Factors in Computing Systems*, 1–16. https://doi.org/10.1145/3411764.3445149

Petrozzino, C. (2021). Who pays for ethical debt in AI?. *AI and Ethics, 1*(3), 205-208.

Raji, I. D., Kumar, I. E., Horowitz, A., & Selbst, A. (2022, June). The Fallacy of AI Functionality. In *2022 ACM Conference on Fairness, Accountability, and Transparency* (pp. 959-972).

Roberts, J., & Lyons, L. (2017). Scoring Qualitative Informal Learning Dialogue: The SQuILD Method for Measuring Museum Learning Talk. Philadelphia, PA: International Society of the Learning Sciences..

Saltz, J., Skirpan, M., Fiesler, C., Gorelick, M., Yeh, T., Heckman, R., ... & Beard, N. (2019). Integrating ethics within machine learning courses. *ACM Transactions on Computing Education (TOCE), 19(4)*, 1-26.

Scott, K. A., Sheridan, K. M., & Clark, K. (2015). Culturally responsive computing: A theory revisited. *Learning, Media and Technology, 40*(4), 412-436.

Shapiro, R. B., Fiebrink, R., & Norvig, P. (2018). How machine learning impacts the undergraduate computing curriculum. *Communications of the ACM, 61*(11), 27-29.

Shapiro, R., & Tissenbaum, M. (2019). New Programming Paradigms. In S. Fincher & A. Robins (Eds.), *The Cambridge Handbook of Computing Education Research* (Cambridge Handbooks in Psychology, pp. 606-636). Cambridge: Cambridge University Press. doi:10.1017/9781108654555.021

Skirpan, M., Beard, N., Bhaduri, S., Fiesler, C., & Yeh, T. (2018, February). Ethics education in context: A case study of novel ethics activities for the CS classroom. In *Proceedings of the 49th ACM Technical Symposium on Computer Science Education* (pp. 940-945).

Strauss, A., & Corbin, J. (1994). Grounded theory methodology: An overview. In N. K. Denzin & Y. S. Lincoln (Eds.), *Handbook of qualitative research* (pp. 273–285). Sage Publications, Inc.

Sulmont, E., Patitsas, E., & Cooperstock, J. R. (2019, February). Can you teach me to machine learn?. In *Proceedings of the 50th ACM Technical Symposium on Computer Science Education* (pp. 948-954).

Sun, B., Da Xu, L., Pei, X., & Li, H. (2003). Scenario-based knowledge representation in case-based reasoning systems. *Expert Systems, 20*(2), 92-99.

Touretzky, D., Gardner-McCune, C., Martin, F., & Seehorn, D. (2019, July). Envisioning AI for K-12: What should every child know about AI?. In *Proceedings of the AAAI conference on artificial intelligence* (Vol. 33, No. 01, pp. 9795-9799).

Wang, N., Lester, J. & Basu, S. (2021). Building Capacity for K-12 Artificial Intelligence Education Research: Workshop 1 Report.

Wild, C. J., & Pfannkuch, M. (1999). Statistical thinking in empirical enquiry. International statistical review, 67(3), 223-248.

Williams, R., Park, H. W., & Breazeal, C. (2019, May). A is for artificial intelligence: the impact of artificial intelligence activities on young children's perceptions of robots. In Proceedings of the 2019 CHI conference on human factors in computing systems (pp. 1-11).

Williams, R., Ali, S., Devasia, N., DiPaola, D., Hong, J., Kaputsos, S. P., ... & Breazeal, C. (2022). AI+ ethics curricula for middle school youth: Lessons learned from three project-based curricula. *International Journal of Artificial Intelligence in Education*, 1-59.

Woodward, Julia, Zari McFadden, Nicole Shiver, Amir Ben-Hayon, Jason C. Yip, and Lisa Anthony. "Using co-design to examine how children conceptualize intelligent interfaces." In *Proceedings of the 2018 CHI Conference on Human Factors in Computing Systems*, pp. 1-14. 2018.

Yip, J., Clegg, T., Bonsignore, E., Gelderblom, H., Rhodes, E., & Druin, A. (2013, June). Brownies or bags-of-stuff? Domain expertise in cooperative inquiry with children. In *Proceedings of the 12th International conference on interaction design and children* (pp. 201-210).

Zimmermann-Niefield, A., Turner, M., Murphy, B., Kane, S. K., & Shapiro, R. B. (2019, June). Youth learning machine learning through building models of athletic moves. In *Proceedings of the 18th ACM international conference on interaction design and children* (pp. 121-132).